\documentclass[aps,floats,subeqn,showpacs,amssymb]{article}
\usepackage{amsmath}
\usepackage{amssymb}
\usepackage{caption}
\usepackage{float}
\usepackage{epstopdf}
\usepackage{comment}
\usepackage{graphicx} % Allows including images
\usepackage{booktabs} % Allows the use of \toprule, \midrule and \bottomrule in tables
\usepackage{perpage} %the perpage package
\usepackage{caption}
\usepackage{float}
\usepackage{titlesec}
\usepackage{nameref}
\usepackage{amsthm}
\usepackage{verbatim}
\usepackage{url}
\usepackage[svgnames]{xcolor}

\newcommand*{\greysquare}{\textcolor{gray}{\blacksquare}}

\newenvironment{myenumerate}{
\begin{itemize}
 \setlength{\itemsep}{1pt}
 \setlength{\parskip}{0pt}
 \setlength{\parsep}{0pt}}{\end{itemize}
}

\theoremstyle{plain}

\theoremstyle{definition}
\newtheorem{defn}{Definition}[section]
\newtheorem{conj}{Conjecture}[section]

\theoremstyle{remark}

\begin{document}

\title{Asymptotic Intrinsic Universality and Reprogrammability by Behavioural Emulation\thanks{Invited chapter contribution to \textit{Advances in Unconventional Computation} by Andrew Adamatzky (ed.), Springer Verlag (forthcoming)}}
\author{
{\bf Hector Zenil\footnote{Corresponding author: hector.zenil AT algorithmicnaturelab.org}} \\ Unit of Computational Medicine, SciLifeLab, Department of\\Medicine Solna, Karolinska Institute, Stockholm, Sweden; \\Department of Computer Science, University of Oxford, UK; and\\
Algorithmic Nature Group, LABORES, Paris, France.\\\\
{\bf J\"urgen Riedel\footnote{jurgen.riedel AT labores.eu}}\\Institut f\"ur Physik, Universit\"{a}t Oldenburg, Germany; and\\
Algorithmic Nature Group, Laboratoire de Recherche\\
Scientifique LABORES, Paris, France.
}
\date{}

\maketitle

\begin{abstract}
We advance a Bayesian concept of \textit{intrinsic asymptotic universality}, taking to its final conclusions previous conceptual and numerical work based upon a concept of a \textit{reprogrammability} test and an investigation of the complex qualitative behaviour of computer programs. Our method may quantify the trust and confidence of the computing capabilities of natural and classical systems, and quantify computers by their degree of reprogrammability. We test the method to provide evidence in favour of a conjecture concerning the computing capabilities of Busy Beaver Turing machines as candidates for Turing universality. The method has recently been used to quantify the number of \textit{intrinsically universal}\index{intrinsic universality} cellular automata\index{cellular automata}, with results that point towards the pervasiveness of universality due to a widespread capacity for emulation. Our method represents an unconventional approach to the classical and seminal concept of Turing universality, and it may be extended and applied in a broader context to natural computation, by (in something like the spirit of the Turing test) observing the behaviour of a system under circumstances where formal proofs of universality are difficult, if not impossible to come by.\\


\end{abstract}

\section{Introduction}

Attempts to answer even the simplest questions about the behaviour of computer programs are bedevilled by uncomputability\index{uncomputability}. The concept of asymptotic intrinsic universality introduced here is based upon a Bayesian approach to emulation by computer programs of other computer programs. The method provides a means to quantify their reprogramming capabilities, associating them with a deciding procedure that asymptotically recognizes computation with a confidence value and sets forth a hierarchy of reprogrammability (see~\cite{reprogrammability2})\index{reprogrammability} based upon the likelihood of a system being, in one degree or another, close to (or removed from) Turing universality.

In~\cite{bbconjecture}, a related conjecture concerning other kinds of simply defined programs was presented, suggesting that all Busy Beaver Turing machines\index{Busy Beaver Turing machines} may be capable of universal computation, as they seem to share some of the informational and complex properties of systems known to be capable of universal computational behaviour. 

We have recently found that most computer programs can be reprogrammed to emulate an increasing number of other (different) computer programs of the same size~\cite{riedelzenil} under a similar \textit{block emulation} transformation or set of compilers of increasing size. We also previously advanced a conceptual framework for reprogrammability based upon the display of different qualitative output behaviours~\cite{reprogrammability1} and modelled as a type of Turing test to determine computational capabilities~\cite{reprogrammability2}. This has been used in connection with an instance of natural computation--in an \textit{in-silico} simulation of\index{Porphyrin molecules} Porphyrin molecules~\cite{natural} in the context of spatial computing\index{spatial computing}. 

Here we advance a Bayesian method, namely \textit{asymptotic intrinsic universality}, that draws everything together and translates the seminal concept of computation universality to degrees of belief and confidence based upon emulation and reprogrammability capabilities applicable to natural computation. We test the method with a case study of the set of Turing machines defined by the Busy Beaver functions.

\section{Methods}

\subsection{The classical Turing machine model}

A Turing machine\index{Turing machine} consists of a finite alphabet set with symbols $\sum=\{0,1,\ldots, k \}$ and states $\{1,2, \ldots n\} \bigcup \{0\}$, with 0 the ``halting state''. The Turing machine ``runs'' on an one-way unbounded tape and for each pair: 
\begin{myenumerate}
\item the machine's current ``state'' $n'$; and
\item the tape symbol $k'$ the machine's head is ``reading''.
\end{myenumerate}

For each pair $(n',k')$ there is a corresponding instruction $(n'',k'',d)$:

\begin{myenumerate}
\item a state $n''$ to transition into (which may be the same as the one it was in). If 0 the machine halts;
\item a unique symbol $k''$ to write on the tape (the machine can overwrite a $1$ on a $0$, a $0$ on a $1$, a $1$ on a $1$, and a $0$ on a $0$), and
\item a direction to move in $d$: $-1$ (left), $1$ (right) or $0$ (none, when halting).
\end{myenumerate}

For $k=2$, there are $(4n + 2)^{2n}$ Turing machines with $n$ states according to this formalism. The output string is taken from the number of contiguous cells on the tape the head has gone through.

\begin{defn}
We denote by $(n,2)$ the set (or space) of all $n$-state 2-symbol Turing machines (with the halting state not included among the $n$ states) and by $T(n,k)$ a specific Turing machine with $n$ states and $k$ symbols.
\end{defn}

\subsection{The Busy Beaver functions}

A \textit{Busy Beaver Turing machine}~\cite{rado} is a Turing machine that, when provided with a blank tape, does a lot of work. Formally, it is an $n$-state $k$-symbol Turing machine started on an initially blank tape that writes a maximum number of 1s or moves the head a maximum number of times upon halting. An online computer program showing the behaviour of these computer programs can be found in~\cite{wolframdemonstration}. 

Most Turing machines never halt, yet Busy Beavers do halt (by definition over the empty tape). We know from algorithmic information theory that among those Turing machines that do halt, most will halt quickly or will perform very little work, yet by definition Busy Beavers are those that perform the greatest amount of work. In a recent investigation~\cite{riedelzenil} focused on cellular automata (CA), we have also shown that most computer programs are candidates for intrinsic universality, and thus for Turing universality.

There are known values for all 2-symbol Busy Beavers up to 4-state Turing machines, and explicit constructions give exact or lower bounds for other state and symbol pairs.

\begin{defn} If $\sigma_T$ is the number of 1s on the tape of a Turing machine $T$ upon halting, then:  $\sum(n)=\max{\{\sigma_T : T\in(n,2) \normalsize{\textbf{ }T(n)\textbf{ }halts}\}}$.
\end{defn}

\begin{defn}
If $t_T$ is the number of steps that a machine $T$ takes upon halting, then $S(n)=\max{\{t_T : T\in(n,2) \normalsize{\textbf{ }T(n)\textbf{ }halts}\}}$.
\end{defn}

$\sum(n)$ and $S(n)$ are noncomputable functions by reduction to the halting problem. Yet values are known for $(n,2)$ with $n \leq 4$.

Busy beavers are the Turing machines that perform more computation among the machines if their same size (by number of states but more appropriately by program length in bits) needed. This follows from Rado's definitions and it means that Busy Beavers have also the greatest Logical Depth, as defined by Bennett~\cite{bennett}. Yet a Busy beaver is required to halt. When running for the longest time or writing the largest number of non-blank symbols, $bb(n)$ has to be clever enough to make wise use of its resources and an instruction away to halt at the end. There is thus evidence that these machines are far from trivial and that for several important measures of complexity they are among most complex, if not the most, yet their computational power is unknown and its investigation would represent a way to connect complexity to computational power. Here we undertake first steps with interesting results.

\subsection{Block emulation and intrinsic universality}

The notion of \textit{intrinsic} computational universality used for cellular automata was an adaptation of classical Turing-universality~\cite{neumann}. \textit{Intrinsic universality} is  stronger than Turing-universality~\cite{ollinger,ollinger2} and the concept can be extended and adapted to other computing systems, including computer programs in general.

\begin{defn}
A computer program of a given size is intrinsically universal if it is able to emulate\index{simulation}\index{emulation} the output behaviour of any other computer program under a coarse-graining\index{coarse-graining} compiler~\cite{ollinger}.
\end{defn}

The so-called \textit{Game of Life} is an example of 2-dimensional cellular automaton that is not only Turing-universal but also intrinsic universal~\cite{durand}. This means that the Game of Life does not only compute any computable function but can also emulate the behavior of any other 2D-dimensional cellular automaton (under rescaling).

\begin{figure}[h!]
\centering 
  \includegraphics[width=120mm]{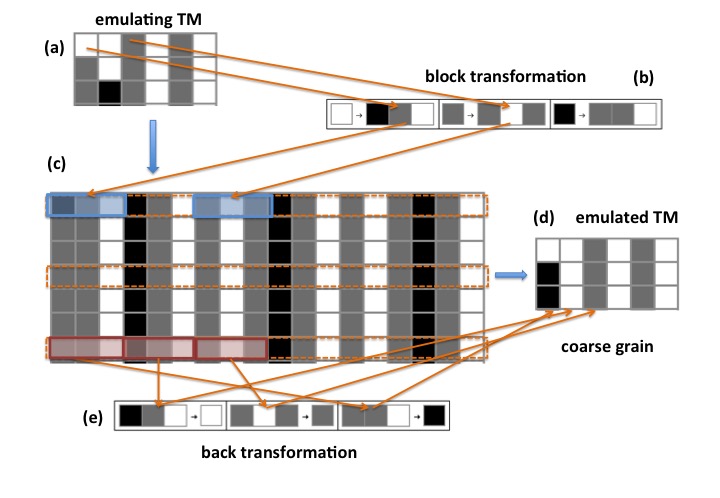}
\caption{\label{fig_3}Illustration of the process of one Turing machine emulating another via a block transformation. In this case (a) shows a $bb(2,3)$ with initial tape $\Box \Box \greysquare \Box \greysquare \Box$ after 2 steps. By performing the block transformation of length 3 (b) on the initial condition of (a), after 6 steps using the same $bb(2,3)$ rule one gets (c). If the output of every 3rd step is taken and the back transformation (d) performed on these outputs, one gets the output (e). This is identical with the output of $TM(2,3)$ with rule number $2\,797\,435$ run on the same initial condition as in (a) for 2 steps. In other words, (e) is the coarse-grained version of the block transformed $bb(4,2)$ (a) which in turn produces the same output as $TM(4,2)$ of rule number $2\,797\,435$. In this picture one cannot see the compiler directly as it is encoded within the internal states of the $bb(4,2)$.
\label{fig_emul_visual}
}
\end{figure}

\begin{defn}
(emulation/simulation by rescaling/coarse-graining): Let $A$ and $B$ be two computer programs. Then $A$ emulates/simulates $B$ if there exists a rescaling/projection $P$ of $A$ such that $f^P_{A} = f_{B}$, where $f_{A}$ and $f_{B}$ are the computed functions of $A$ and $B$. We consider $P$ a compiler to translate $A$ into $B$ (see Fig.~\ref{fig_3}).
\end{defn}

The exploration of the computing capabilities of computer programs can then proceed by \textit{block emulation}, whereby the scale of space-time diagrams of a computation are found and rescaled/coarse-grained.

The emulations here explored are related to an even stronger form of \textit{intrinsic universality}, namely linear-time intrinsic universality~\cite{ollinger}, which implies that all emulations carry only a linear overhead as a result of our brute force exploration of the compiler and rule space. This is because the coarse-graining emulation is of a block of fixed length and therefore what one can consider a compiler (another computer program of fixed size).

Following these ideas, one can try out different possible compilers\index{compilers} and see what type of computer programs a specific computer program is able to emulate. The \textit{linear} block transformation was suggested in~\cite{StephenWolfram1983, StephenWolfram1986}.

\subsection{Turing machine emulation}

The exploration of the emulating space of Turing machines (TM) is more complicated than for Cellular Automata because the space-time diagram does not contain the head configuration state of the Turing machine. 

We ran the random TMs and the Busy Beaver Turing machines for the number of steps given by $S(n)$. For example, for $n=4$ states, $S(n)=107$, given by the Busy Beaver $bb(4)$. We looked for all transformations which allow a back transformation for block sizes 2 to 4 and only considered (2-symbol, 4-state) and 3-symbol, 2-state) Busy Beaver Turing machines and a sample of random Turing machines of the same size.

\begin{figure}[h!]
\centering 
  \includegraphics[width=125mm]{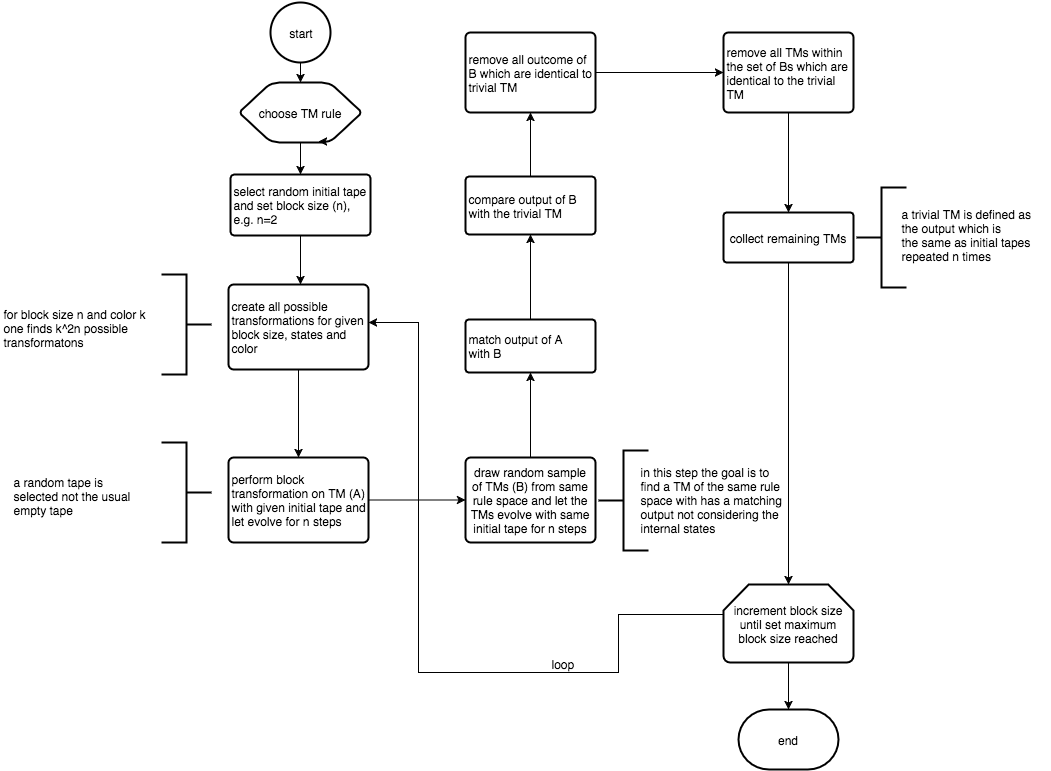} \\
\caption{\label{diagram} Flow diagram of emulation of TMs.}
\end{figure}

To ascertain which TM from the same rule space corresponded to the emulated Busy Beaver or TM, we adopted the following algorithm: 

For a $n$-state and $k$-symbol Turing machine (TM), we enumerated all possible block transformations $P(n,k)$ of given block size $n$ ($n$-tuples), e.g. $P(2,3)=\Box \rightarrow \blacksquare \greysquare, \blacksquare \rightarrow \greysquare \Box, \greysquare \rightarrow \greysquare \greysquare$ for a 3-symbol, 2 state TM. We found a total of $k^{2n}$ possible transformations. We applied each member of the set of possible transformations to a TM of the corresponding rule space, in this paper that of a Busy Beaver ($bb$) or a randomly selected TM given a randomly initialized tape. We then let the TM evolve for n steps. Then we  took every n output line of the TM and performed a back transformation on the output, e.g. $P(2,3)^{-1}= \blacksquare \greysquare \rightarrow \Box, \greysquare \Box \rightarrow \blacksquare, \greysquare \greysquare \rightarrow \greysquare $. At the same time we drew a TM of the same rule space out of a random sample and let it evolve for $n$ steps using the same initial tape. If the output was a valid output of a TM, we tried to match it with the output of the Busy Beaver or random TM described above. In order to exclude trivial emulations, we filtered out all those emulated TMs which are just a $n$-time repetition of the initial tape. It is important to note that we are not taking the initial states of the TMs into account. We are just focusing on the output of TMs when performing the block transformations.

\subsubsection{Busy Beaver conjectures}

These facts suggest the following conjectures, which are also relevant to the dynamic behaviour of a set of simply-described machines characterized by universal behaviour. 

In previous work we explored these conjectures relating to Busy Beavers with numerical approximations of their sensitivity to initial conditions~\cite{bbconjecture} and the qualitative behaviour that initial conditions induce over space-time diagrams~\cite{zenilchaos}. Which was similar to work we did on the Game of Life~\cite{gameoflife}. 

\begin{defn}
A \textit{weak} universal Turing machine is a machine that allows its initial tape to be in a non all-`blank' configuration.
\end{defn}

If $bb(n)$ is weak universal, then it is allowed to start either from a periodic tape configuration or an infinite sequence produced by e.g. a finite automaton. In other words, these are machines that are Turing universal not necessarily running on non-empty tapes.

\begin{conj}
\label{conjecture}
The Busy Beaver conjecture(s) as advanced in~\cite{bbconjecture} establish(es) that:
\begin{myenumerate}
\item (strong version): For all $n>2$, $bb(n)$ is Turing universal.
\item (sparse version): For some $n>2$, $bb(n)$ is Turing universal.
\item (weak version): For all $n>2$, $bb(n)$ is weak Turing universal.
\item (weakest version): For some $n>2$, $bb(n)$ is (weak) Turing universal.
\end{myenumerate}
\end{conj}

Here we provide evidence in favour of all conjectures in the form of an increasingly monotonic asymptotic intrinsic universal behaviour.

It is known that among all 2-state 2-symbol Turing machines, none can be universal. $bb(n)$, as defined by Rado~\cite{rado}, is a Turing machine with $n$ states plus the halting state. $bb(2)$ is thus actually $bb(2,3)$, a 3-state 2-symbol machine in which one state is specially reserved for halting only. If $bb$ is unary, then it will be assumed to be a 2-symbol Turing machine, otherwise it will be denoted by $bb(n,k)$.

\begin{figure}[ht!]
{
\centering 
  \includegraphics[width=80mm]{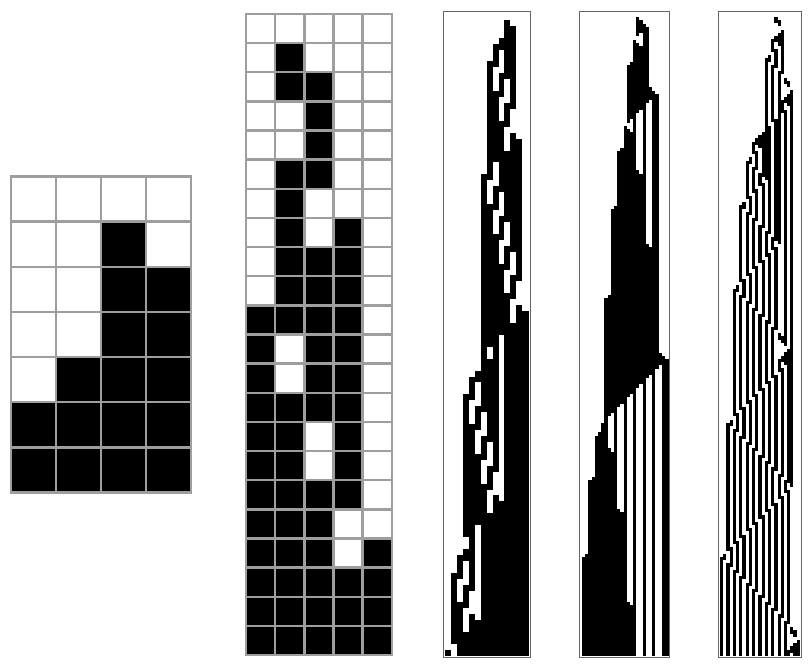}  \includegraphics[width=40mm]{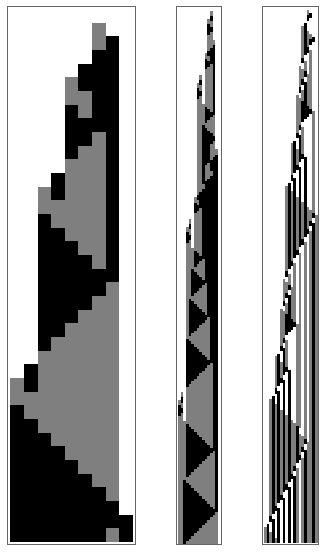}
\caption{\label{fig_1} Typical space-time evolution/behaviour of \textit{Busy Beaver Turing machines}. The first 6 figures from left to right correspond to Busy Beaver machines with 2-symbols and 2 to 6 states (for illustration purposes only those $<4$ were plotted with a background mesh) for which the first 3 have exact ($S(n)$) runtime values (6, 21, 107). For the rest a cutoff value was arbitrarily chosen, so as to provide an optimally effective illustration. The behaviour of a Busy Beaver cannot be a trivial repetition because it does have to avoid getting into an infinite cycle in order to halt.}
}
\end{figure}

\section{Results}

\subsection{A Bayesian\index{Bayesian method} approach to Turing universality}

We looked into the number of compilers up to a certain size for which a computer program can emulate other computer programs of the same size (e.g. in number of states for TMs, number neighbours for CAs, or description bits in general). Given all the unknown priors and the uncertainty in the degree of belief, we need a basic  function that:

\begin{myenumerate}
\item Is increasingly monotonic. Normalizing by total number of explored compilers should provide a measure for comparison, but the function itself should only count the number of emulations.
\item $f(x)>0$ when $x>0$. Evidently any emulation should amount to a non-zero value.
\item Nonlinearly converges to 1. We want a function that ``slowly'' converges to a positive value and
\item Incorporates a degree of belief weighting the number of emulations found.
\end{myenumerate}

Because intrinsic universality implies Turing universality~\cite{ollinger}, this approach is of relevance in finding the reprogramming capabilities of classical and unconventional computing systems.

\begin{figure}[ht!]
{
\centering
\includegraphics[width=90mm]{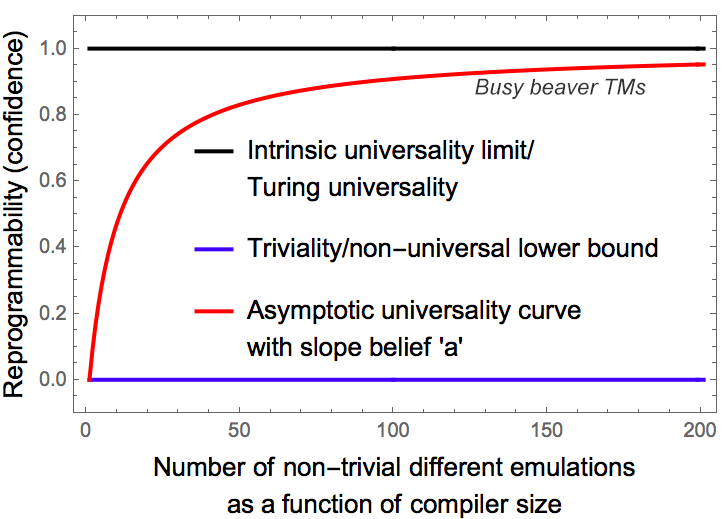}
\caption{\label{universality} Asymptotic intrinsic universality curve ($ax/(ax+1)$, made continuous for illustration purposes) is the Bayesian approach to the otherwise seminal but abstract concept of computation universality applicable to both abstract and natural/unconventional computation. For example, we found evidence in favour of a conjecture postulating that Busy Beaver Turing machines are somewhere on the asymptotic universality curve, highly so if the degree of belief according to $a$ assigns it a higher confidence every time that such a machine in question is able to emulate some other.}
}
\end{figure}

The exact shape of the function has no essential meaning as long as it is concave and complies with the above requirements. A canonical function is $ax/(ax+1)$, where $x \in \mathbb{N}^+$ is the number of different non-trivial emulations of a system under evaluation and $a \in (0,1]$ the degree of belief modifying the rate of convergence, in this case $a=1$ (see Fig.~\ref{universality}). We then define the asymptotic intrinsic universality of a computing system $s$ as:

\begin{defn}(asymptotic intrinsic universality): Let $s$ be a computing machine of fixed size and $x$ the number of non-equivalent (e.g. under coarse-graining) emulations of other computing systems of the same size that $s$ can emulate, then  $\Delta(s)=ax/(ax+1)$ is the function that retrieves a confidence value of reprogrammability based upon the intrinsic universality of $s$ according to belief $a$.
\end{defn}

\subsection{Case study: Busy Beaver functions}

Here we provide evidence in favour of the Busy Beaver conjectures by way of the different qualitative behavioural properties they display and their intrinsic universality capabilities.

\subsubsection{Qualitative behaviour analysis}

Among the intuitions suggesting the truth of one of these conjectures, is that it is easier to find a machine capable of halting and performing unbounded computations for a Turing machine if the machine already halts after performing a sophisticated calculation, than it is to find a machine showing sophisticated behaviour whose previous characteristic was simply to halt. This claim can actually be quantified, given that the number of Turing machines that halt after $t=n$ for increasing values of $n$ decreases exponentially \cite{calude,delahaye,zeniltheoremprover}. In other words, if a machine capable of halting is chosen by chance, there is an exponentially increasing chance of finding that it will halt sooner rather than later, meaning that most of these machines will behave trivially because they will not have enough time to do anything interesting before halting\index{halting problem}.

\begin{figure}[ht!]
{
\centering 
A \hspace{5cm} B\\
  \includegraphics[width=50mm]{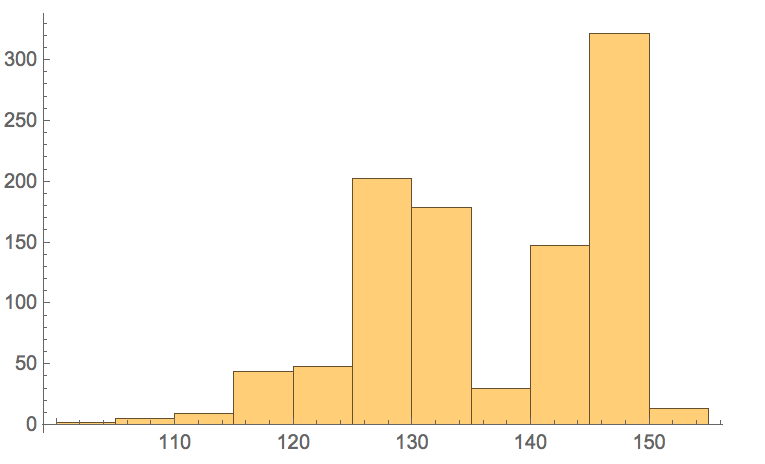} \hspace{1cm} \includegraphics[width=50mm]{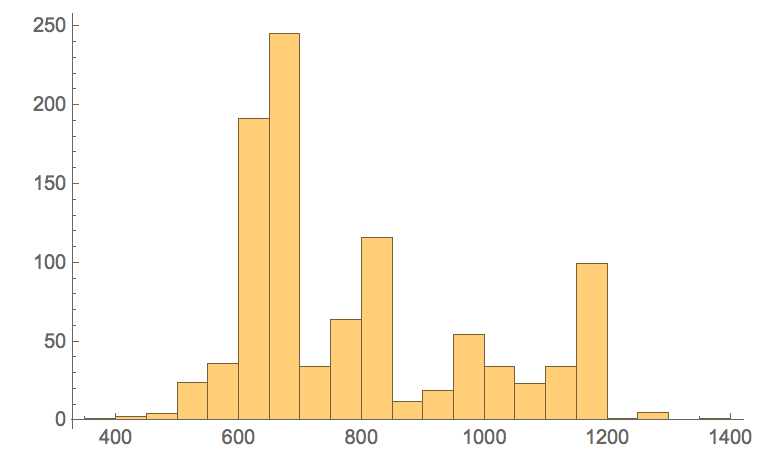}\\
  \medskip
C \hspace{5cm} D\\
    \includegraphics[width=50mm]{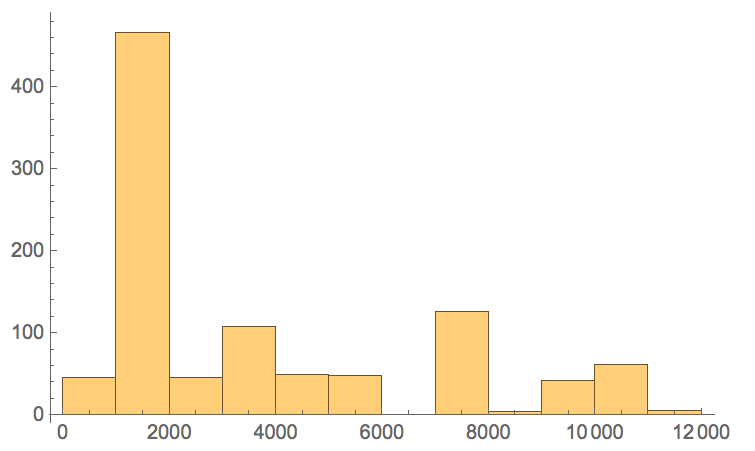} \hspace{1cm}
      \includegraphics[width=50mm]{3_1000.png}\\
\medskip
E \hspace{5cm} F\\
      \includegraphics[width=45mm]{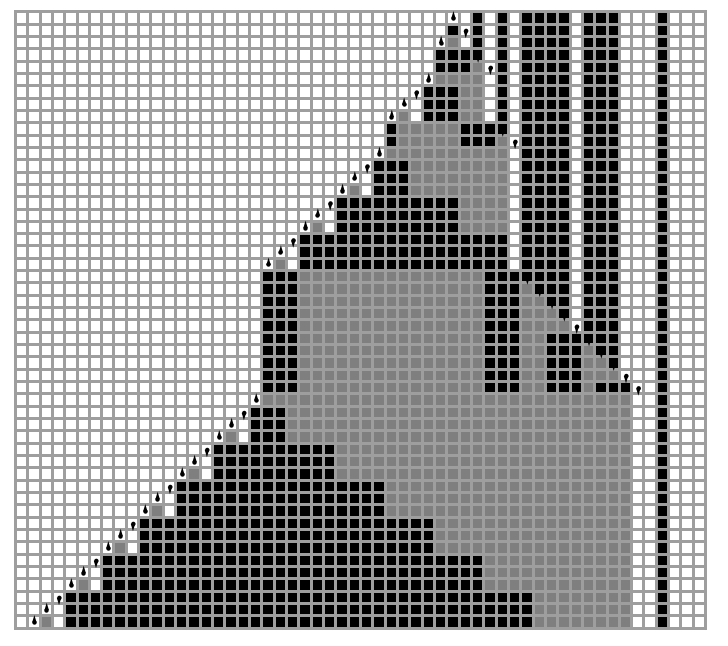}\hspace{.8cm}
      \includegraphics[width=55mm]{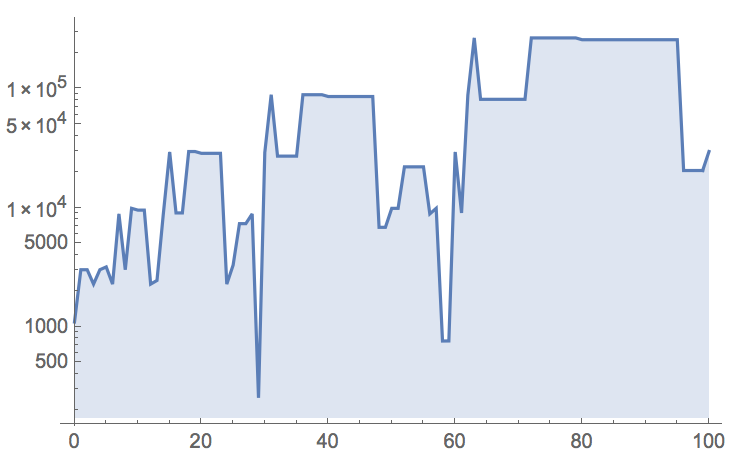}\\
\caption{\label{mainfig} A-D: Histograms of the compressed lengths ($x$ axis using Compress) of the space-time diagrams of $bb(n)$ for $n=$3 to 6 for $1\times 10^3$ steps each, showing accumulation of different qualitative behaviours. E: A right-left compressed behaviour of a Busy Beaver runnning for $1.5 \times 10^3$. Only rows for which the head has moved further to the right or left than ever before are kept, a method suggested in~\cite{wolfram}. F: Function computed by the Busy Beaver $b(5,2)$ for consecutive initial conditions 1 to 100 in binary.}
}
\end{figure}

\begin{figure}[ht!]
{
\centering 
    \hspace{-.2cm}  \includegraphics[width=55mm]{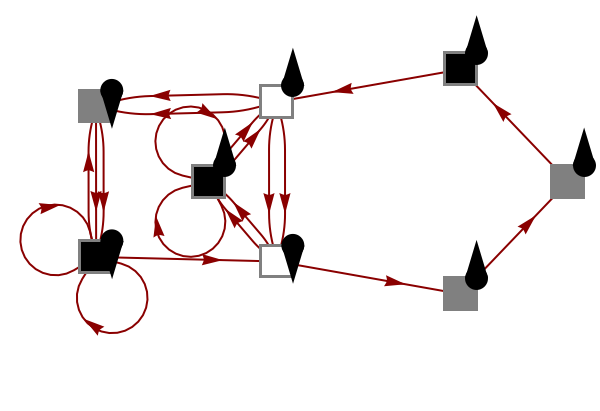}\hspace{.5cm}
      \hspace{.5cm}\includegraphics[width=20mm]{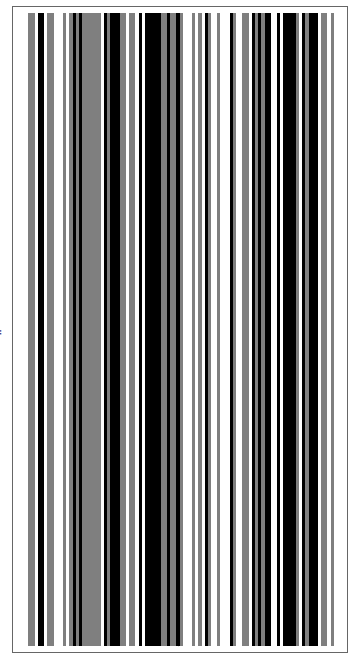} \hspace{.3cm} \includegraphics[width=19.5mm]{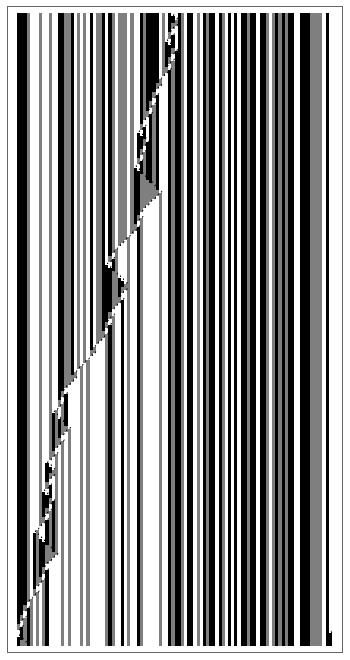}
\caption{\label{mainfig2} Left: State diagram after 20 steps (state 1 is a down-tick, state 2 is an up-tick). Right: Two runs from different ``random'' initial conditions of length 100 bits showing (left) a quick halting (computation of the identity) and (right) an apparently random movement of the head for another initial condition running on the same 4-state 3-symbol Busy Beaver Turing machine.}
}
\end{figure}

Fig.~\ref{mainfig} provides a summation of the behavioural investigation of Busy Beaver machines. Histograms show the different qualitative behaviour in bimodal and multimodal discrete distributions. The multimodality is not an effect of the size of the initial condition that grows smoothly by $\log(n)$, nor of the stepwise behaviour of the lossless compression algorithm (Compress based upon Deflate). If it were an effect of the length of the initial condition, then Subfigs. B-D would look like Subfig. A, which is not the case. They display genuinely different behaviours (see Fig.~\ref{mainfig2}(right)). 

The state diagram in Fig.~\ref{mainfig2}(left) suggests how to choose an initial configuration for the machine to enter into an infinite loop (e.g. connected cycle on the left), and therefore how to enter into a never-halting computation, a requirement for (weak) Turing universality. Fig.~\ref{mainfig2}(right) shows the behaviour of $bb(4,3)$ for 2 different initial conditions, one for which it halts (or ``computes'' the identity) and another for which the computation goes on in a rather complex head movement fashion.

\subsubsection{Reprogrammability of Busy Beavers by block emulation}

\begin{figure}[h!]
\centering 
  \includegraphics[width=54mm]{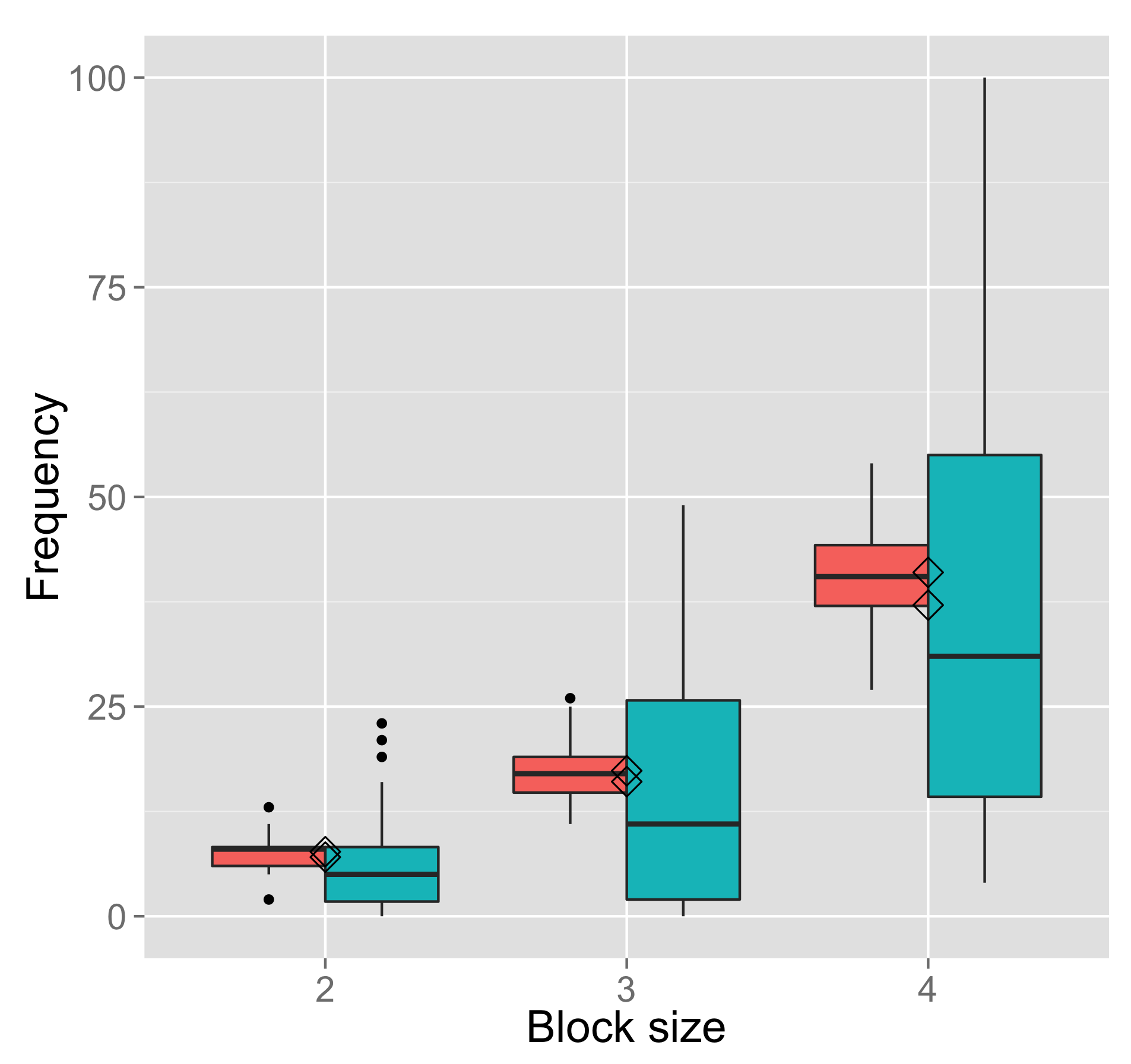} \includegraphics[width=66mm]{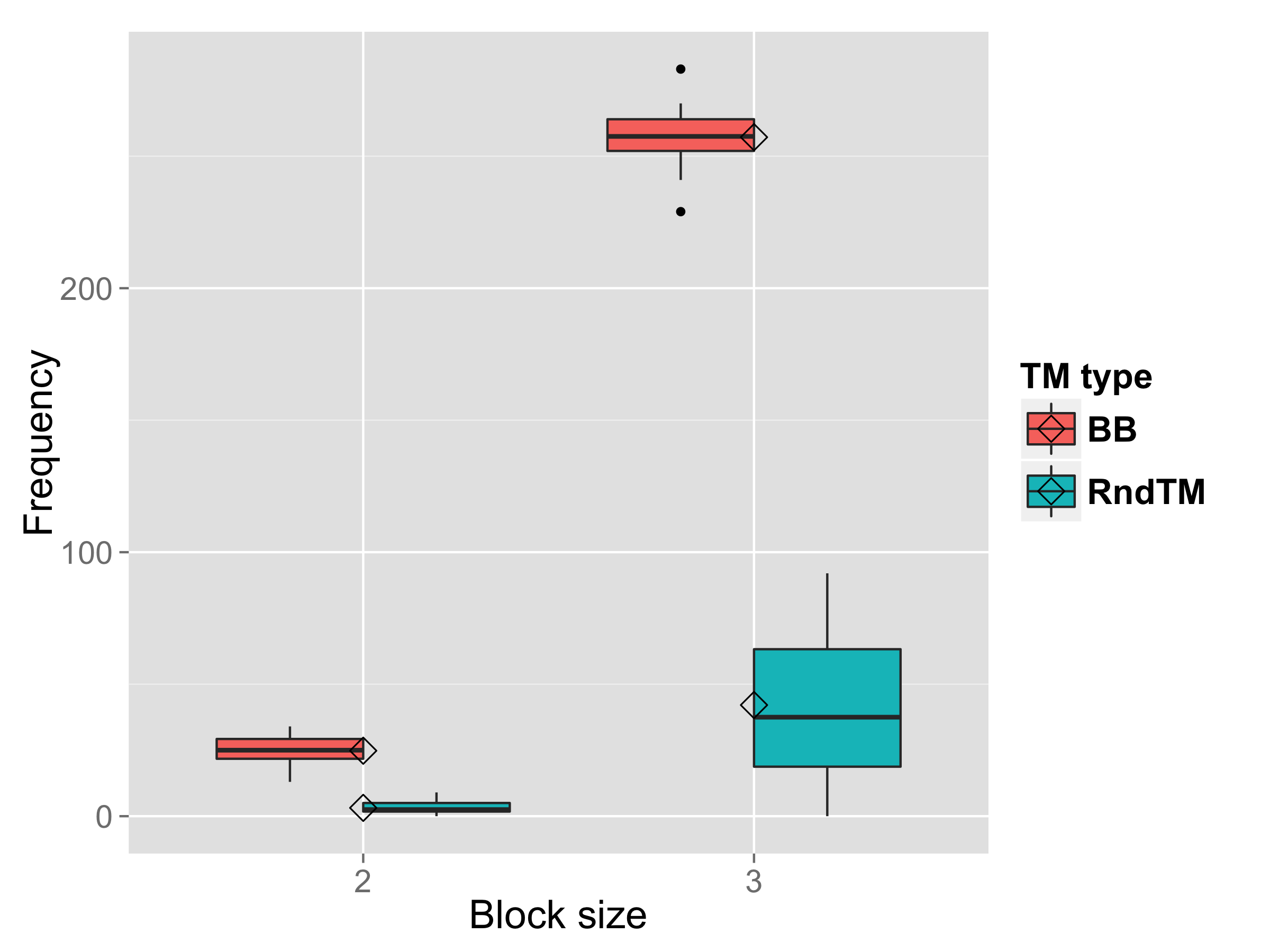}\\
    \includegraphics[width=54mm]{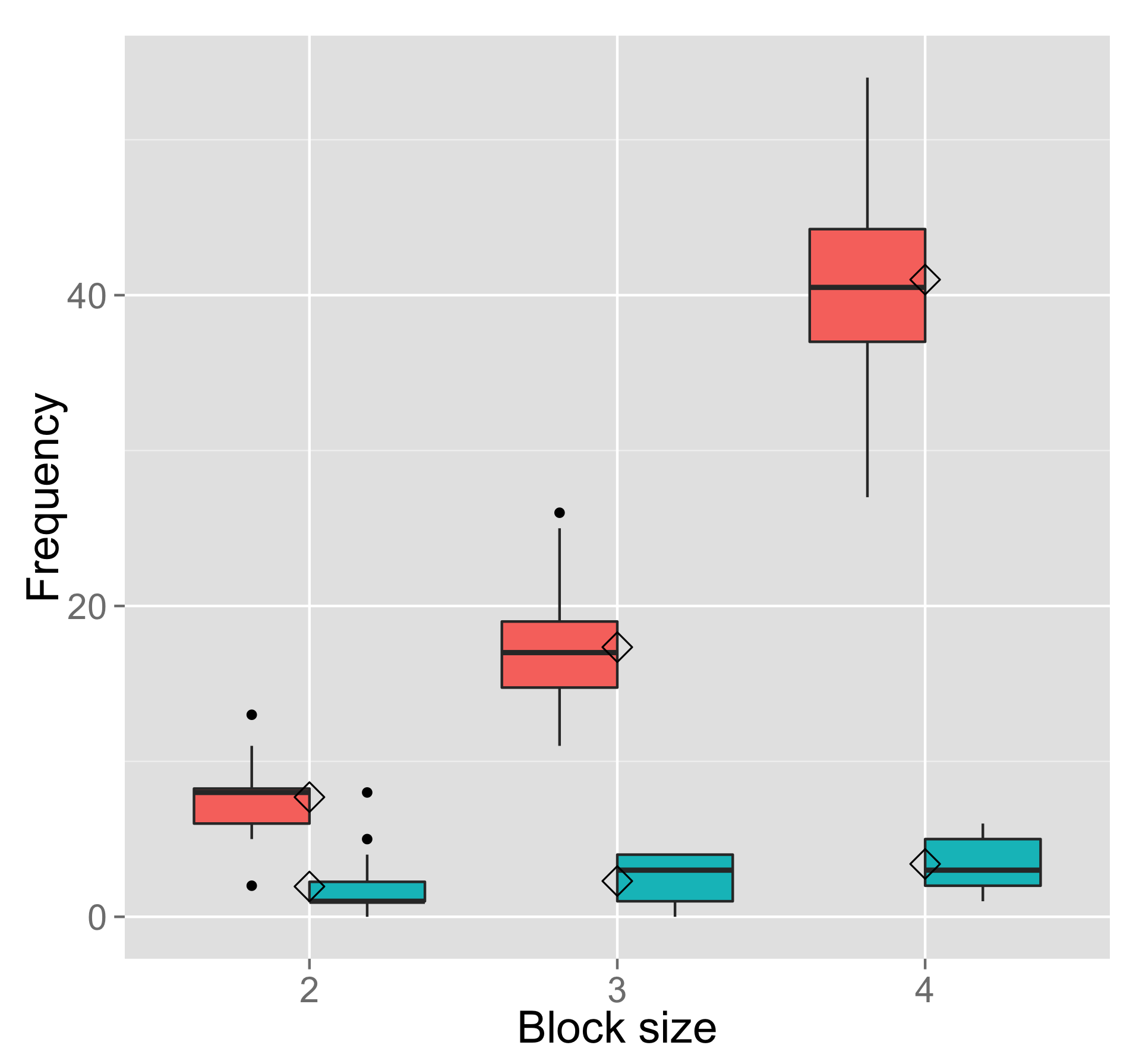} \includegraphics[width=66mm]{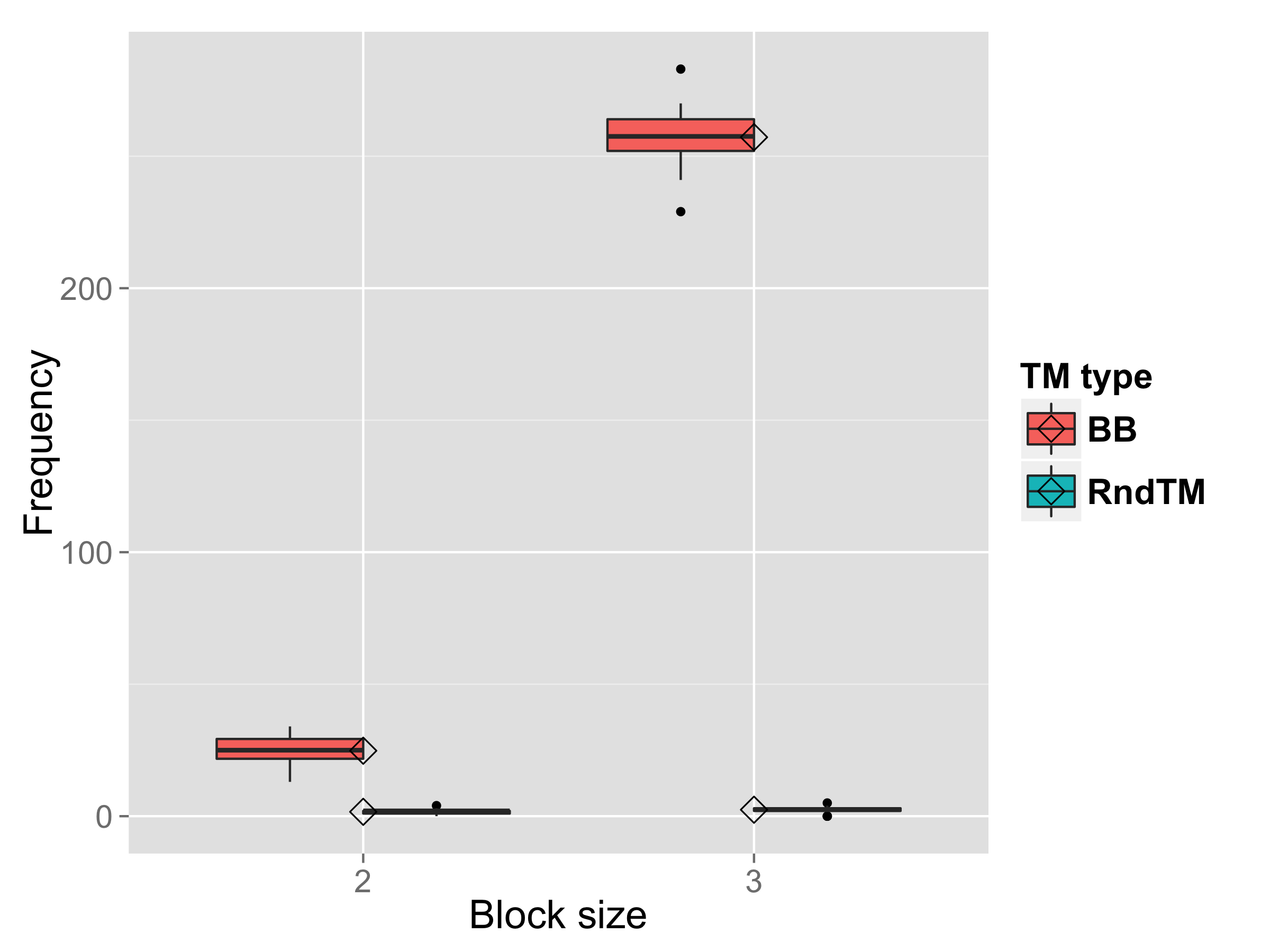}
\caption{\label{fig_2} Top: Boxplots showing the differences in the emulation power of (left) $bb(4,2)$ versus a set of randomly selected TMs in $(4,2)$, and (right) $bb(2,3)$ versus a set of randomly selected TMs in $(2,3)$. The data show how many emulations on average a set of Busy Beavers of a given rule space and a set of random TMs selected from the same rule space can produce for given block sizes. The data shows a variance for both TM types, since the output of valid block transformations is compared with the output of a TM sample taken from the same rule space. Trivial TMs (c.f. flow chart in Fig~\ref{diagram}) are excluded. Each emulation is counted, even if it corresponds to the same TM. The diamond shapes represent the mean of the data points. Bottom: Same plots, but only TM evolutions with different hash values (from their evolution) are counted, i.e. only distinct TMs are counted, rendering the difference between Busy Beavers and random Turing machines even more prominent.}
\end{figure}

Fig.~\ref{fig_2} shows that Busy Beavers are much more capable of emulating the behaviour of other (non-trivial) Turing machines than the control case, a sample of random Turing machines from the same rule space size (i.e. all machines are of the same size). This is consonant with theoretical expectations~\cite{calude}.

\subsubsection{Busy Beavers are candidates for Turing universality}

The capacity for universal behaviour implies that a system is capable of being reprogrammed and is therefore reactive to external input. It is no surprise that universal systems should be capable of responding to their input and doing so succinctly, if the systems in question are efficient universal systems. If the system is incapable of reacting to any input or if the output is predictable (decidable) for any input, the system cannot be universal. 

We have here provided evidence that Busy Beavers comply with all the requirements for Turing universality and must therefore be considered a very interesting non-trivial set of Turing machines that are candidates for Turing universality. 

Evidence in favour of the conjectures is based upon the following observations:

\begin{myenumerate}
\item Busy beavers produce space-time diagrams of the highest complexity compared to the space-time diagrams of other rules in the same rule space.
\item Busy beavers show qualitatively different behaviour for different initial conditions; they can halt and it is not difficult to devise ways to perform non-halting computations based upon infinite loops, especially for non-empty inputs.
\item The small set of Busy Beavers investigated emulate a larger number of other (non-trivial) Turing machines on average compared to random Turing machines of the same size. In other words, we found evidence indicating that $\Delta(bb(n,k))>\Delta(RndTM(n,k))$, where $RndTM(n,k) \in bb^c$ is a random Turing machine in the complement set of the Busy Beavers $bb^c$, and the confidence level $a$ is fixed.
\item Thus the measure of asymptotic intrinsic universality that we defined $\Delta(bb(n,k))$ converges to 1 much faster than $\Delta(RndTM(n,k))$.
\end{myenumerate}

Asymptotic intrinsic universality is strictly stronger than Turing universality. Fig.~\ref{fig_2}: Random TM statistics serve as a control because we know that the set of machines that either quickly halt or never halt are of density measure 1, and will therefore end up dominating the average emulation with $\Delta(RndTM(n,k))/nk \sim 0$ for $n,k \rightarrow \infty$. So if we find, as we in fact did, that $\Delta(bb(n,k))/nk$ grows faster than $\Delta(RndTM(n,k))/nk$, we would be demonstrating with a high level of confidence that Busy Beaver Turing machines have greater reprogramming capabilities and are candidates for intrinsic universality, and therefore Turing universality.

\section{Discussion}

%It may be objected that there is some tension between what we have used to theoretically justify the control experiments relating to the emulation capabilities of Busy Beaver Turing machines, and what we claim to have found here and elsewhere as regards the pervasiveness of Turing universality. However, this tension is only apparent, because the number of non-universal computer programs is very low for a fixed input, e.g. empty, as in the theoretical results provided by Calude et al.~\cite{calude} Nonetheless, when the domain of operation of these machines is constrained to a subset of new possible initial conditions (effectively a space of \textit{coarse-graining compilers}), there is a qualitative change in the expected behaviour of the same machines, behaviour that is not without a trade-off (compilers are difficult to find and become sparser under at least two conditions~\cite{riedelzenil}). In other words, the theoretical results hold for unrestricted machine domains but not for restricted ones (under the non-trivial compiler space).\index{compiler space}

\subsection{Universality versus reprogrammability in natural computation}

We have brought together several concepts that are relevant and applicable to natural computation where, e.g., resources are often scarce and computation occurs independently of the substrate, making for concepts that are disembodied, independent not only of specific hardware but of models and formalisms (e.g. whether one can define a halting configuration). 

On the one hand, there is the use of the concept of intrinsic universality, which our definition of asymptotic universality relies upon. Intrinsic universality as originally formulated for cellular automata does not require a halting configuration. This makes it applicable to natural computation, because a halting state is an arbitrary choice--the option to design a state as a halting one--which is meaningless in natural computation. Furthermore, the coarse-graining only takes into consideration the output configuration rather than the state configuration, which is consistent with recent extensions for \textit{membrane computing}\index{membrane computing} or P systems~\cite{gheorghe}\index{P systems}, where a computation with only one possible output can be reached through many different paths, regardless of the internal states transited through en route. Indeed, since we are not looking `inside' the TM (its internal states), we are treating it as a black box\index{black box computation} (see Fig.~\ref{fig_3}) on which we perform an external observer test. The compiler used to look at the internal states is a behaviourally shallow one. Interestingly, the transformed TM (in Fig.~\ref{fig_3}, a Busy Beaver) does lock immediately into the same pattern. Again, one would need to visualize the internal states to see a difference between other emulations producing the same output. 

On the other hand, in a world where ``emptiness'' or simple/completely regular initial conditions cannot be guaranteed, weak universality is more realistic. The concept of asymptotic universality is based upon and adapted to deal with these situations in the context of natural computation where a system may be a black box but its behaviour can be reinterpreted (by emulation) and exploited. Of course one difficulty is to identify different behaviours in order to undertake a behavioural comparison, and this is why we have also introduced complexity indices that can serve as tools to quantify the space-time evolutions of systems or their representations.\index{natural computation}

A limitation of any empirical approach is that of a non-universal system able to emulate an increasing number of non-universal systems may not be universal but an open question is whether there is a threshold N above which a system is universal after simulating N number of other systems. This is also why the emulation results should be complemented by an analysis of the complexity of the emulations themselves. While we do know that, for example, finite automata are bounded by the kind of complexity they can produce by the complexity of the set of regular languages, the connection between these qualitative differences and the computational power of the systems in this context is a future subject of further investigation. 

The chief advantage of this approach is the amenability to non-formal evidence of the reprogrammability of less conventional systems, where formal proofs of universality are difficult, if not impossible to come by. In other words, while the method does disclose universal systems at the limit, it does not rule out non-universal ones, thus producing possible false positives. However limited, any false positive is still a  reprogrammable system, thereby providing a more natural/pragmatic definition of \textit{natural universality}.

\subsection{The Busy Beaver conjectures}

It would not have been possible to anticipate that the behaviour displayed would have been that of Busy Beavers, despite their complexity for empty inputs. Nor could the low emulation capabilities of all other trivial and non-trivial machines in the complement set of the Busy Beavers $bb^c$ have been anticipated, because they are no longer being tested and quantified over the full set of possible initial conditions but over the subset that allow the emulation of other computer programs (Turing machines) of the same (growing) size. In other words, what we are exploring is the Cartesian product $P \times C$ of the pairs $(p,c)$, where $p\in P$ is a computer program (e.g. a Turing machine) and $c\in C$ a compiler that maps $p$ onto $p'\in P$ of size $|p'|=|p|$ (in this case the number of states, but in the general the number of bits, i.e. its\index{Kolmogorov complexity} Kolmogorov complexity~\cite{chaitinbb}). 

Here we explored the reprogrammable space, a subset of the the space of all computer programs for either a specific input or, equivalently (per Turing universality), for all inputs. This also means that most of the machines that either halt almost immediately and therefore do nothing interesting, or else never halt, can actually be effectively reprogrammed, and the results obtained here and in~\cite{zenilchaos,riedelzenil} strongly suggest that they may even be candidates for intrinsic universality (i.e. the ability to emulate any other computer program under a coarse-graining compiler),\index{coarse-graining computation} a stronger concept than that of Turing universality.

\section{Conclusion}

The set of Busy Beaver machines describes an (enumerable) infinite set of Turing machines characterized by a particular specific behaviour. If the conjectures are true according to the evidence we have provided, the result is more surprising, because a describable property determines the computational power of this non-trivial infinite set of Turing machines. Here we have taken these ideas a step further in the direction of an empirical proposal for considering statistical computational evidence of computational universality. Because of the undecidability of the halting problem we may never obtain stronger evidence of the computational capabilities of these computer programs.

We have introduced a novel experimental and methodological Bayesian approach to theoretical computing challenges that circumvents traditional limitations imposed by classical definitions, in particular related to undecidability, unreachability and universality and deals with pragmatic unconventional reprogramming by behavioural emulation rather than through attempting producing formal analytical proofs, which are not only difficult, but impossible in general, specially in the realm of natural computation where we think these new concepts and methods are more relevant.

\end{document}